\documentclass[submission,copyright,creativecommons]{eptcs}
\providecommand{\event}{HCVS 2022} 
\usepackage{breakurl}             
\usepackage[dvipsnames]{xcolor}
\usepackage{tikz}
\usepackage{listings}
\usepackage{parcolumns}
\usepackage{amsmath}
\usepackage{paralist}

\usepackage{listings}

\lstdefinestyle{floating}{%
    xleftmargin=10pt,%
    xrightmargin=5pt,%
    aboveskip=4mm,%
    belowskip=4mm,%
    fontadjust=true,%
    columns=[c]flexible,%
    keepspaces=true,%
    basewidth={0.5em, 0.425em},%
    tabsize=2,%
    basicstyle=\ttfamily,%
    commentstyle=\rm,%
    keywordstyle=\bfseries,%
    mathescape=true,%
    captionpos=b,%
    framerule=0.3pt,%
    firstnumber=0,%
    numbersep=1.5mm,%
    numberstyle=\tiny,%
    float=tbp,%
    frame=tblr,%
    framesep=5pt,%
    framexleftmargin=3pt,%
    abovecaptionskip=\smallskipamount,%
    belowcaptionskip=\smallskipamount,%
} 

\usetikzlibrary{arrows}
\usetikzlibrary{shapes}
\usetikzlibrary{patterns}
\usetikzlibrary{backgrounds}
\usetikzlibrary{positioning}
\usetikzlibrary{calc}

\tikzstyle{switch}  = [circle,draw=black, thick, fill=yellow!80!green,inner sep=0.4pt]

\newcommand{\ClauseSet}{\mathit{HC}}
\newcommand{\Leq}{\sqsubseteq}
\newcommand{\Plus}{$\mathbin{\tikz [x=1.4ex,y=1.4ex,line width=.2ex, black] \draw (0.5,0) -- (0.5,1) (0,0.5) -- (1,0.5);}$}%

\newtheorem{definition}{Definition}
\newtheorem{example}{Example}

\title{OptiRica: Towards an Efficient Optimizing Horn Solver}
\author{Hossein Hojjat
\institute{Tehran Institute for Advanced Studies, Khatam University\\ University of Tehran, Iran
}
\email{hojjat@ut.ac.ir}
\and
Philipp R\"ummer
\institute{University of Regensburg, Germany\\Uppsala University, Sweden}
\email{\quad philipp.ruemmer@it.uu.se}
}
\def\titlerunning{OptiRica: Towards an Efficient Optimizing Horn Solver}
\def\authorrunning{H. Hojjat \& P. R\"{u}mmer}
\begin{document}
\maketitle

\begin{abstract}
  This paper describes an ongoing effort to develop an optimizing
  version of the Eldarica Horn solver. The work starts from the
  observation that many kinds of optimization problems, and in
  particular the MaxSAT/SMT problem, can be seen as search problems on
  lattices. The paper presents a Scala library providing a
  domain-specific language (DSL) to uniformly model optimization
  problems of this kind, by defining, manipulating, and systematically
  exploring lattices with associated objective functions. The
  framework can be instantiated to obtain an optimizing Horn
  solver. As an illustration, the application of an optimizing solver
  for repairing software-defined networks is described.
\end{abstract}

\section{Introduction}

Constrained Horn Clauses have proven to sit at a sweet spot: the language of Horn clauses is expressive enough to capture interesting properties of complex systems; at the same time, their mathematical properties (in particular the existence of least models) enable efficient algorithms and solvers that scale to real-world applications. Nevertheless, for a long time, people have tried to extend the language of Horn clauses, while keeping some of its convenient properties. Among others, it was proposed to extend Horn clauses with well-foundedness predicates~\cite{BjornerGMR15}, universally quantified literals in the clause body~\cite{BjornerMR13}, existentially quantified literals in the clause head~\cite{BeyenePR13}, and disjunctions in heads~\cite{Beyene2015TemporalPV}.

This paper presents ongoing work in a similar direction: the development of optimizing solvers in which side conditions are formulated in terms of Horn clauses, and optimization objectives are characterized using finite lattices. This setting naturally generalizes MaxSAT/SMT~\cite{maxsat} (and minimal unsatisfiability) to the setting of Horn clauses, and thus captures many forms of analysis and reasoning tasks; for instance, the exploration of all counterexamples of a set of clauses, the inference of safe parameters or sufficient pre-conditions, or program repair.

Our work builds on a lattice-based optimization framework designed for
the purpose of interpolant
exploration~\cite{DBLP:journals/acta/LerouxRS16} and network
repair~\cite{HojjatRMCF16}. In those earlier papers, only a
prototypical implementation was provided that was not directly
reusable in other contexts. This paper recapitulates the optimization
framework, gives an overview of ongoing implementation work in the
context of the Horn solver Eldarica~\cite{Eldarica18}, and illustrates
the use of optimization for repair. We include several code examples
showing how our new optimization library can be used to formulate and
solve optimization problems.

\paragraph{Related work.}

In addition to the research already mentioned, our work is related to
the computation of maximum specifications of
functions~\cite{DBLP:conf/popl/AlbarghouthiDG16}, and weakest
solutions of Horn clauses~\cite{DBLP:conf/pldi/SFMD21}. Other
approaches proceed in a counterexample-guided manner, and refine
solutions until a maximum or weakest solution has been
found~\cite{DBLP:conf/popl/AlbarghouthiDG16,DBLP:conf/pldi/SFMD21}. The
methodology behind our framework is different, as we require an
upfront specification of the optimization objective and search space
in the form of a finite lattice; our work is closer in spirit to
MaxSAT/SMT~\cite{maxsat}.

\section{The Lattice-Based Optimization Framework}

\subsection{The Framework}

The lattice-based optimization framework~\cite{HojjatRMCF16,DBLP:journals/acta/LerouxRS16} is inspired by the MaxSAT/SMT paradigm~\cite{maxsat}, but generalizes MaxSAT in two ways: (i)~where MaxSAT can be seen as a search on the powerset lattice induced by some set of constraints, we consider arbitrary finite lattices; and (ii)~where side conditions in MaxSAT are given in terms of SAT or SMT constraints, the feasibility of solutions in the lattice optimization framework can be defined by any computable function.

The central definition needed to explain the framework is the notion of an \emph{optimization lattice;} the following definition is a slightly generalized version of the definition in earlier papers~\cite{HojjatRMCF16}:
\begin{definition}[Optimization lattice]
  \label{def:optLattice}
  An \emph{optimization
    lattice} is a triple~$(\langle L, \Leq_L \rangle, F, \mathit{obj})$ consisting
  of a complete lattice $\langle L, \Leq_L \rangle$, a downward-closed feasibility predicate $F \subseteq L$, and a monotonically increasing objective function~$\mathit{obj} : L \to D$ to a set $D$ that is totally ordered.
\end{definition}
We call the elements in $F$ \emph{feasible.} Note that the
predecessors of feasible elements are also feasible, and the
successors of infeasible elements are also infeasible.  An
element~$l \in L$ is \emph{maximal feasible} if $l$ is feasible, but
all of its successors are infeasible.

\begin{definition}
  An element~$l_{\mathit{max}} \in L$ is called \emph{optimal} if it is
  maximal feasible, and it is the case that
  $\mathit{obj}(l_{\mathit{max}}) = \max \{ \mathit{obj}(l) \mid l \in
  F\}$.
\end{definition}

Examples of useful optimization lattices are powerset lattices,
interval lattices, as well as lattice
products~\cite{HojjatRMCF16}. Algorithms to compute maximal feasible
and optimal elements are able to handle large finite lattices (e.g.,
lattices with $>10^{10}$ elements) by representing those lattices
symbolically~\cite{HojjatRMCF16,DBLP:journals/acta/LerouxRS16}. Those
algorithms are based on three main principles: greedy optimization,
which is achieved by walking upward in a lattice until reaching a
maximal feasible element; the computation of incomparable elements, to
identify the next starting point in the lattice after discovering one
maximal feasible element; and the inference of upper bounds on
feasible elements in the lattice for early pruning.  The computation
of incomparable elements is related to hitting set methods used in
MaxSAT~\cite{maxsat}, while upper feasibility bounds have similarities
with the use of blocking clauses in SAT.

\subsection{Implementation}

\begin{figure}[tb]
  \includegraphics[width=\linewidth,trim=5 270 240 5]{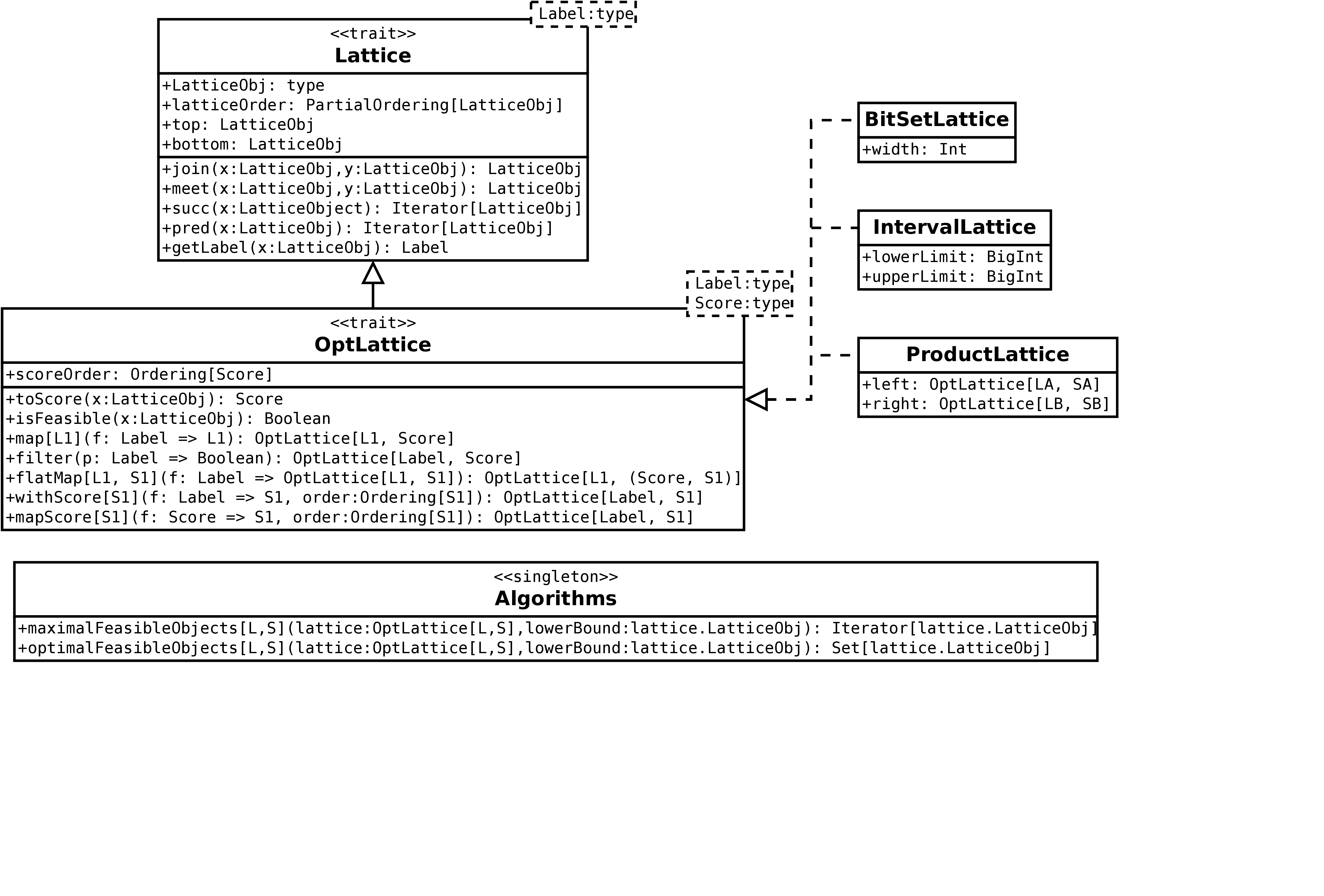}
  
  \caption{Class diagram of lattice types}
  \label{fig:latticeImpl}
\end{figure}

We are in the process of developing a Scala library for lattice-based
optimization.\footnote{\url{https://github.com/uuverifiers/lattice-optimiser}}
At the moment, our library does not include a parser for a modeling
language for expressing optimization problems; instead, we define an
embedded domain-specific language (EDSL) using a set of basic
lattice types, and operators to derive new lattices from those basic
lattices.

An overview of the lattice classes is provided in
Figure~\ref{fig:latticeImpl}. In general, a lattice is symbolically
described through a type~\texttt{LatticeObj} of the nodes in the
lattice, a partial order~\texttt{latticeOrder} on those nodes, and
methods for computing joins, meets, successors, and predecessors of
nodes. Moreover, lattices are labeled, with a method~\texttt{getLabel}
mapping lattice nodes to elements of some defined label type. While
\texttt{Label} is a type parameter of the \texttt{Lattice} trait, the
type \texttt{LatticeObj} is an abstract type member, and should be
seen as an existential type: every lattice is associated with some
type of lattice nodes, but given a lattice no assumptions can in
general be made about the node type.

Optimization lattices are derived from general lattices, but in
addition provide an objective function~\texttt{toScore} for mapping
lattice nodes to some \texttt{Score} type, a feasibility
predicate~\texttt{isFeasible}, as well as a monadic interface
consisting of \texttt{map}, \texttt{filter}, and \texttt{flatMap}
methods. The \texttt{map} method is used to redefine the labeling of a
lattice, by specifying a function from the old to a new label type;
\texttt{filter} strengthens the feasibility predicate of a lattice by
conjoining a new constraint; and \texttt{flatMap} is used to construct
product lattices by mapping labels to optimization lattices. The
objective function~\texttt{toScore} can be redefined using two
methods: the method~\texttt{withScore} creates lattices in which the
score of a node is computed as a function of the node label; and
\texttt{mapScore} mutates the score of each node by applying some
function~\texttt{f} to it.

The functions to compute optimal elements in lattices are collected in
the class~\texttt{Algorithms}. The
method~\texttt{maximalFeasibleObjects} enumerates the maximal feasible
nodes above some \texttt{lowerBound} in a given lattice;
method~\texttt{optimalFeasibleObjects} returns the set of maximal
feasible nodes above \texttt{lowerBound} with maximum score.

\begin{example}  
  A simple example of an optimization problem expressed using the
  library is given in Listing~\ref{lst:squares}. The code
  snippet\footnote{Complete working example in~
    \url{https://github.com/uuverifiers/lattice-optimiser/blob/master/src/test/scala/lattopt/SquareTests.scala}}
  solves the problem of computing the subset of the
  numbers~$\{0, 1, \ldots, 16\}$ that
\begin{inparaenum}[(i)]
\item has the property that it does not contain any number~$x$ as well
  as its square~$x^2$; and
\item has the maximum element sum.
\end{inparaenum}
To model the problem, in line~4 the powerset lattice of the set
$\{0, 1, \ldots, 16\}$ is declared; powerset lattices are an instance
of the \texttt{BitsetLattice} class. The objective function is defined
in line~5 to be the sum of the elements of a set, and the feasibility
condition is defined in line~6. The sets of maximal feasible and
optimal solutions are computed in lines~9 and 11, respectively. Since
the methods to compute solutions are randomized, in line~0 the used
random number generator is initialized.
\end{example}

\begin{lstlisting}[language=Scala,float=tb,caption={An optimization problem over a powerset lattice.},
                   numbers=left,frame=tlrb,label=lst:squares]
implicit val randomData = new SeededRandomDataSource(123)

// The powerset of the set {0, ..., 16}; subsets that contain
// the square of any of their elements are infeasible
val latt = PowerSetLattice(0 to 16).
             withScore(_.sum).
             filter { s => !(s exists { x => s contains x*x }) }

// there are four maximal feasible objects
println(Algorithms.maximalFeasibleObjects(latt)(latt.bottom).size)

// and the optimum, maximizing the sum of its elements:
// {2, 5, 6, 7, 8, 9, 10, 11, 12, 13, 14, 15, 16}
println(Algorithms.optimalFeasibleObjects(latt)(latt.bottom))
\end{lstlisting}

\section{Lattice-Based Optimization Modulo Horn Constraints}

\subsection{Parameterized Clauses}

The optimization framework can immediately be used to derive solvers for \emph{MaxCHC,} the maximum satisfiability problem over Horn clauses. Given a set $\ClauseSet$ of constrained Horn clauses, consider the powerset lattice~${\cal P}(\ClauseSet)$, as well as the feasibility predicate~$F = \{ S \subseteq \ClauseSet \mid S \text{~is satisfiable} \}$ that can be implemented using an existing Horn solver. Appropriate timeouts have to be used when checking the satisfiability of subsets of $\ClauseSet$.
Solutions of MaxCHC are then the maximal feasible elements of the induced optimization lattice. Partial and weighted MaxCHC (including both hard and soft constraints, or giving weights to clauses, respectively) can be modeled by choosing an appropriate objective function~$\mathit{obj}$.

To compute minimally unsatisfiable sets (MUSes), we can construct the
inverted (or dual) powerset lattice induced by some set of clauses, and
the feasibility predicate
$F = \{ S \subseteq \ClauseSet \mid S \text{~is unsatisfiable} \}$.

The challenge, of course, is to make such constructions efficient, since the exploration of the optimization lattice will require many expensive satisfiability checks on subsets of $\ClauseSet$. A practical implementation therefore requires an underlying \emph{incremental} Horn solver that is tailored to the case of solving many similar queries. We are in the process of extending our solver Eldarica~\cite{Eldarica18} for this purpose, using the following notion of parameterized clauses:
\begin{definition}[Parameterized Clause Set]
  A \emph{parameterized clause} is a constrained Horn clause over a set $R = R_{\mathit{sym}} \cup R_{\mathit{par}}$ of relation symbols, including a subset~$R_{\mathit{par}}$ of relation symbols called \emph{parameters}. If $m : R_{\mathit{par}} \to \mathit{Constr}$ is a function mapping every $n$-ary parameter to a formula/constraint over $n$ free variables, then the \emph{instance}~$\ClauseSet[m]$ of a  parameterized clause set~$\ClauseSet$ is obtained by substituting every parameter~$p \in R_{\mathit{par}}$ with the constraint~$m(p)$.
\end{definition}
The incremental version of Eldarica is able to directly process
parameterized clauses, and postpone the instantiation of parameters
as late as possible, thus minimizing the
amount of work that has to be repeated when solving different clause
set instances. Eldarica can also reuse counterexamples and
solutions. As a future extension, we plan to add functionality to
reapply CEGAR predicates across instances.

This yields the following recipe for defining optimization problems
modulo Horn clauses: (i)~define a set~$\ClauseSet$ of parameterized
Horn clauses, such that the instances of the clauses cover the
intended search space; (ii)~define a
lattice~$\langle L, \Leq_L \rangle$ representing the search space;
each element~$o \in L$ is labelled with a substitution~$m_o$;
(iii)~use one of the feasibility
predicates~$F_{\mathit{sat}} = \{ o \in L \mid \ClauseSet[m_o]
\text{~is satisfiable} \}$ or
~$F_{\mathit{unsat}} = \{ o \in L \mid \ClauseSet[m_o] \text{~is unsatisfiable}
\}$, implemented using an incremental Horn solver; (iv)~choose a
suitable objective function on the lattice. The downward-closedness of
the feasibility predicate can be ensured syntactically, for instance
by choosing a monotonic labeling function~$m_o$ on the lattice, and
restricting parameters~$R_{\mathit{par}}$ to the clause bodies.

\begin{lstlisting}[language=Scala,float=tb,caption={Definition of parameterized clauses, and the resulting optimization problem.},
  numbers=left,frame=tlrb,label=lst:paramClauses]
import [...]
SimpleAPI.withProver { p =>
  import p._
  
  val Inv  = createRelation("I", Seq(Integer, Integer, Integer))
  val Flag = for (i <- 0 to 3) yield createRelation("f" + i, Seq())

  val Seq(x, y, n) = createConstants(3)

  val clauses = List(
    Inv(0, 0, n)         :- (              n >  0,   Flag(0)()),
    Inv(x + 1, y + 1, n) :- (Inv(x, y, n), y <  n,   Flag(1)()),
    Inv(x + 2, y + 1, n) :- (Inv(x, y, n), y <  n,   Flag(2)()),
    false                :- (Inv(x, y, n), x >= 2*n, Flag(3)())
  )

  def set2map(s : Set[Predicate]) =
    (for (f <- Flag) yield (f -> if (s(f)) TRUE else FALSE)).toMap

  val l1 = PowerSetLattice(Flag).withScore([...])).map(set2map(_))
  val l2 = ClauseSatLattice(l1, clauses, Flag.toSet)

  println(Algorithms.optimalFeasibleObjects(l2)(l2.bottom))
}
\end{lstlisting}

\subsection{Implementation}\label{sec:maxCHC}

We are developing the integration of the lattice optimization library
with Eldarica as a separate library
OptiRica.\footnote{\url{https://github.com/uuverifiers/optirica}}
Parameterized clauses can in this setting either be created
programmatically, or be read from an SMT-LIB or Prolog file using the
front-end of Eldarica. In
Listing~\ref{lst:paramClauses},\footnote{Complete working example in
  \url{https://github.com/uuverifiers/optirica/blob/main/src/test/scala/optirica/ClauseSatTest.scala}}
the former approach is chosen, and four clauses are defined over the
relation symbols $R_{\mathit{sym}} = \{\mathtt{Inv}\}$ and parameters
$R_{\mathit{par}} = \{\mathtt{Flag}(0), \ldots, \mathtt{Flag}(3)\}$
(lines~4--14). Each of the four clauses is labeled with a
parameter~$\mathtt{Flag}(i)$, which can be set to $\mathit{false}$ to
disable some of the clauses.

The system consisting of all four clauses is unsatisfiable. We can
therefore solve a MaxCHC problem, and compute maximum satisfiable
subsets of the four clauses. The optimization lattice for this problem
has to provide the substitutions~$m_o$ instantiating parameters with
formulas.  For this, in line~19 the powerset lattice of the set of
flags is created, equipped with some suitable objective function, and
then labeled with the functions computed by \texttt{set2map}. In
line~20, the class~\texttt{ClauseSatLattice} is used to define those
lattice nodes as feasible for which the set of instantiated clauses is
satisfiable. \texttt{ClauseSatLattice} receives as arguments the
lattice~\texttt{l1} to be filtered, the parameterized
clauses~\texttt{clauses}, and the set~\texttt{Flag} of parameters.

Line~20 has the same effect as a direct call to the
lattice~\texttt{filter} function,
\begin{lstlisting}[language=Scala,mathescape]
val l2 = l1 filter { m => $\mathit{isSat}($clauses[m]$)$ }$,$
\end{lstlisting}
but \texttt{ClauseSatLattice} includes the optimizations discussed
above, such as applying the Horn solver Eldarica incrementally,
caching results, and reusing solutions and counterexamples. The
constructed lattice~\texttt{l2} has three maximal feasible elements,
corresponding to disabling the first, third, or fourth clause.

The OptiRica library also provides a class
\texttt{ClauseUnsatLattice}, which declares those lattice nodes as
feasible for which some instantiated set of clauses is unsatisfiable.
In combination with an inverted powerset lattice
(\texttt{PowerSet.inverted(...)}), this functionality can be used to
compute MUSes.

\section{A Case Study: Repair in Software-defined Networking}

\subsection{Overview}

\begin{figure}[b]
\noindent\begin{minipage}[b]{0.43\linewidth}
    \centering
\begin{tikzpicture}[node distance = 1cm, auto]
\node [] (H1) {$H_1$} ;
\node [switch, above=0.2cm of H1] (T1) {$T_1$} ;
\node [left of=H1] {\scriptsize Host};
\node [switch, above of=T1] (A1) {$A_1$} ;
\node [left of=A1] {\scriptsize Aggregation};
\node [switch, right of=T1] (T2) {$T_2$} ;
\node [left of=T1] {\scriptsize ToR};
\node [switch, right of=T2] (T3) {$T_3$} ;
\node [switch, right of=T3] (T4) {$T_4$} ;
\node [switch, above of=T3] (A3) {$A_3$} ;
\node [switch, above of=T4] (A4) {$A_4$} ;
\node [below =0.2cm of T2] (H2) {$H_2$} ;
\node [below =0.2cm of T3] (H3) {$H_3$} ;
\node [below =0.2cm of T4] (H4) {$H_4$} ;
\node [below =0cm of H4, align=left] () {Not safe\\~ for $H_1$.};
\node [switch, above of=T2] (A2) {$A_2$} ;
\coordinate (Middle-A1-A2) at ($(A1)!0.5!(A2)$);
\coordinate (Middle-A3-A4) at ($(A3)!0.5!(A4)$);
\node [switch, above of=Middle-A1-A2] (C1) {$C_1$} ;
\node [left=0.8cm of C1] {\scriptsize Core};
\node [switch, above of=Middle-A3-A4] (C2) {$C_2$} ;

\node (M) [node distance=0.5cm,below right of=A3] {\Plus};


\path [draw,->] (H1) -- (T1);
\path [draw] (H2) -- (T2);
\path [draw] (H3) -- (T3);
\path [draw] (H4) -- (T4);
\path [draw] (T1) -- (A2);
\path [draw] (T2) -- (A2);
\path [draw] (T2) -- (A1);
\path [draw] (T1) -- (A1);
\path [draw] (A1) -- (C1);
\path [draw] (A2) -- (C1);
\path [draw] (A3) -- (T3);
\path [draw] (A3) -- (T4);
\path [draw] (A4) -- (T3);
\path [draw] (A4) -- (T4);
\path [draw] (C1) -- (A4);
\path [draw] (C1) -- (A3);
\node[right=0.1cm of C1]{\footnotesize \textit{filter}($H_1$)};
\path [draw,dashed] (C2) -- (A1);
\path [draw,dashed] (C2) -- (A3);
\path [draw,dashed] (C2) -- (A4);
\path [draw,dashed] (C2) -- (A2);

\end{tikzpicture}

\caption{Software-defined network~\cite{HojjatRMCF16}}
\label{fig:example}
\end{minipage}
\hfill
\begin{minipage}[b]{0.57\linewidth}
\centering
\begin{tikzpicture}[thick, scale=0.9]
    \draw 
		node(RB) at (0,.5){$(-\infty,2]$}
		node(MB) at (1.5,1.5){$[2,2]$}
		node(MC) at (1.5,-.5){$(-\infty,3]$}
		node[draw,ellipse,inner sep=0.7pt,fill=gray!50,pattern=north west lines, pattern color=gray!40] (YC) at (3,.5){$[2,3]$}
		node(YD) at (3,-1.5){$(-\infty,4]$}
		node(YB) at (4,2.5){$\emptyset$};
    \draw[shift={(0,-.5)}]
		node(NB) at (4,2){$[3,3]$}
		node(NC) at (4,0){$[2,4]$}
		node(ND) at (4,-2){$(-\infty,+\infty)$}
		node(BB) at (5,1){$[3,4]$}
		node(BC) at (5,-1){$[2,+\infty)$}
		node(TA) at (6.5,2){$[4,4]$}
		node(TB) at (6.5,0){$[3,+\infty)$}
		node[draw,ellipse,inner sep=0.7pt,fill=gray!50,pattern=north west lines, pattern color=gray!40](H) at (8,1) {$[4,+\infty)$};
  \foreach\xa/\xb in{RB/MB,RB/MC,MB/YB,TA/YB,MB/YC,MC/YC,MC/YD,%
                                       YB/NB,YC/NB,YC/NC,YD/NC,YD/ND,
                                       NB/BB,NC/BB,NC/BC,ND/BC,BB/TA,BB/TB,BC/TB,TA/H,TB/H}
      {\draw[-](\xa)--(\xb);}
  \node at (5.3,-2.2) {\rotatebox{50}{$\sqsubseteq$}};
\end{tikzpicture}

\caption{Inverted interval lattice~\cite{HojjatRMCF16}}
\label{fig:lattice}
\end{minipage}

\end{figure}

We have used an earlier, tailor-made implementation of an optimizing Horn solver in the domain of repairing SDN configurations~\cite{HojjatRMCF16}, and outline now how this case study can be mapped to our generic framework. Consider Figure~\ref{fig:example}, which shows a three-layer topology in data centers.
The network sends the packets originating from a host upward and then back downward to the destination host.  
Assume the host $H_1$ sends traffic to $H_2$ and $H_3$, but this traffic should not reach $H_4$.
To implement this policy, the operator installs, e.g., a forwarding rule at $C_1$
to filter packets from $H_1$ going towards $A_4$ and also disable the link $A_3{-}T_4$
for good measure (in the figure, this disabled link is indicated by a
``\Plus'' symbol.)

For maintenance, the network operator has turned off the core switch $C_2$.
When the network operator brings back $C_2$, it causes a safety violation, 
since there is a new path from $H_1$ to $H_4$.
There are multiple repair solutions to this violation: the repair engine may disconnect the links $A_1{-}C_2$ and $A_2{-}C_2$, 
or take $C_2$ offline and return the network to its initial state, or 
rewrite the traffic from $H_1$ to another type of
traffic by modifying packet headers, or add filters for $H_1$ traffic 
on a number of links: $\{A_1{-}C_2, A_2{-}C_2\}$, $\{C_2{-}A_4,
A_4{-}T_4\}$, $\{A_4{-}T_4\}$, etc.

To derive such possible repair strategies, we assume that the safety of the network has been modelled as a set of Horn clauses~\cite{HojjatRMCF16}. This set includes clauses representing the topology and the configuration of the network, for instance will the sending of packets from $A_4$ to $T_4$ be modelled as a clause~$s_{T_4}(\mathit{pkt}', \mathit{trc}') \leftarrow s_{A_4}(\mathit{pkt}, \mathit{trc}) \wedge \phi$. Intuitively, a predicate $s_{T_4}(\mathit{pkt}, \mathit{trc})$ expresses that packets~$\mathit{pkt}$ can reach $T_4$ via path $\mathit{trc}$.
To search for repairs that involve filtering between $A_4$ and $T_4$, the clause can be replaced with a parameterized clause~$s_{T_4}(\mathit{pkt}', \mathit{trc}') \leftarrow s_{A_4}(\mathit{pkt}, \mathit{trc}) \wedge \mathit{filter}(\mathit{pkt}) \wedge \phi$. The relation symbol~$\mathit{filter} \in R_{\mathit{par}}$ is the parameter, and will be substituted with concrete constraints during optimization. To search for repairs that will block packets to a certain port range, an interval lattice like in Figure~\ref{fig:lattice} can be chosen, and the substitutions be defined as $m_{[l, u]}(\mathit{filter}) = (\mathit{pkt}.\mathit{port} \not\in [l, u])$. Maximal feasible elements in the lattice represent safe configurations in which a minimal range of ports is blocked; for instance, the intervals~$(2, 3)$ and $[4, +\infty)$. To model a larger space of possible repairs, multiple parameters can be introduced, and a product optimization lattice be chosen. To express the preference of certain kinds of repairs, an appropriate objective function can be added.

\begin{lstlisting}[language=Scala,float=tb,caption={Derivation of destination filters~\texttt{dstFilter} and type filters~\texttt{typFilter} for the link $A_4{-}T_4$.},
  numbers=left,frame=tlrb,label=lst:filtering]
val dstFilter = createRelation("dstFilter", Seq(Sort.Integer))
val typFilter = createRelation("typFilter", Seq(Sort.Integer))

val clauses = List(
  [...]
  t4(dst,typ) :- (a4(dst,typ), dst === 4,
                  dstFilter(dst), typFilter(typ))),
  [...]
)

// Possible dst filters. v(0) refers to the argument of dstFilter
val dstLatt = for (s <- PowerSetLattice(1 to 4))
              yield disjFor(for (t <- s) yield v(0) === t)
    
// Possible typ filters. v(0) refers to the argument of typFilter
val typLatt = for (s <- PowerSetLattice(0 to 7))
              yield disjFor(for (t <- s) yield v(0) === t)
    
val latt1 =   for (c1 <- dstLatt; c2 <- typLatt)
              yield Map(dstFilter -> c1, typFilter -> c2)
val latt2 =   latt1 mapScore { p => p._1 + p._2 }
val latt3 =   ClauseSatLattice(latt2, clauses,
                               Set(dstFilter, typFilter))
  
println(Algorithms.optimalFeasibleObjects(latt3)(latt3.bottom))
\end{lstlisting}

\subsection{Implementation}

A simple model of the
network\footnote{\url{https://github.com/uuverifiers/optirica/blob/main/src/test/scala/optirica/NetworkTest.scala}}
includes four clauses defining network ingress, 40~clauses for the
network, and property clauses (e.g., $H_1$ traffic not reaching
$H_4$). This model enables us to compute different kinds of repairs to
eliminate the path from $H_1$ to $H_4$. Using a MaxCHC encoding, as in
Section~\ref{sec:maxCHC}, it can be derived that there are 12
different minimal sets of network clauses (i.e., maximal feasible
lattice nodes) that can be removed to re-establish a safe network; two
of those only require to disable a single link, namely either
$A_4{-}T_4$ or $C_2{-}A_4$.

The derivation of more fine-grained filters is illustrated in
Listing~\ref{lst:filtering}. We show a simplified setting in which
predicates only consider the destination and type of packets; type~$0$
is associated with packets originating from $H_1$. To infer filters
for the link~$A_4{-}T_4$, we rewrite the corresponding clause to
include predicates~\texttt{dstFilter} and \texttt{typFilter} in its
body (lines~5--6), and define lattices describing the possible choices
of filtering. In lines~11--12, a destination filter is defined by
choosing a subset~$s$ of the set~$\{1, \ldots, 4\}$, and then
constructing the formula~$\bigvee_{i \in s} x = i$. In lines~15--16,
similarly filters on type are defined. Lines~18--19 create the product
of the two lattices and define the resulting substitution. The
objective function is defined to be the total number of destinations
and types that are not filtered out in line~20.  This optimization problem has
two optimal solutions, namely to filter out either packets of type~0
or packets with destination~$H_4$ on link $A_4{-}T_4$.

\section{Conclusions}

We have outlined work towards optimizing Horn solvers. While the
presented research is still work in progress, we believe that the
lattice-optimization libraries are already useful tools at this point.
Next steps in this project include the implementation of further types
of lattices, an improved, more efficient interface between the
library and the Horn solver, and a more polished language (e.g.,
SMT-LIB-based) for making optimization available to users.

\paragraph{Acknowledgments.}

We would like to thank the reviewers for helpful comments.
Philipp R\"ummer is supported by the Swedish Research Council (VR) under
grant~2018-04727, by the Swedish Foundation for Strategic Research
(SSF) under the project WebSec (Ref.\ RIT17-0011), by the Wallenberg
project UPDATE, and by grants from Microsoft and Amazon Web Services.

\bibliographystyle{eptcs}
\bibliography{generic}
\end{document}